\documentstyle[fleqn,epsf]{article}
\newcommand{\AmS}{{\protect\the\textfont2
  A\kern-.1667em\lower.5ex\hbox{M}\kern-.125emS}}
\newcommand{\bD}{{\bf D}}
\newcommand{\calO}{{\cal O}}
\newcommand{\psibar}{\overline{\psi}}
\newcommand{\ihalf}{\mbox{\small $\frac{i}{2}$}}

\def\mbf{\bf}        

\newcommand{\plus}{\makebox[15pt][c]{$+$}}
\newcommand{\minus}{\makebox[15pt][c]{$-$}}
 
\newcommand{\er}[2]
{\hskip-0.5em\raisebox{0.08em}{\scriptsize{$\;\begin{array}{@{}l@{}}
\plus\makebox[0.25em][r]{#1\hfill} \\[-0.12em]
\minus\makebox[0.25em][r]{#2\hfill} 
\end{array}$}}}
\newcommand{\err}[2]
{\hskip-0.5em\raisebox{0.08em}{\scriptsize{$\;\begin{array}{@{}l@{}}
\plus\makebox[0.50em][r]{#1\hfill} \\[-0.24em]
\minus\makebox[0.50em][r]{#2\hfill} 
\end{array}$}}}

\title{$B$ and $D$ Meson Decay Constants in Lattice QCD
}

\author{ 
 Aida X. El-Khadra,$^a$  Andreas S. Kronfeld,$^b$ \\
Paul B. Mackenzie,$^b$ Sin\'ead M.~Ryan,$^b$ 
and James N. Simone$^{a}$ \\[0.2cm]
\it $^a$ Dept. of Physics, University of Illinois, Urbana, Illinois  61801\\
\it $^b$ Fermilab, PO Box 500,  Batavia, Illinois 60510 
}

\begin{document}

\maketitle

\begin{abstract}
We have calculated the decay constants of $B$ and $D$ mesons
with lattice QCD.
We use an $\calO (a)$ improved  action that takes
 light quark actions as a starting point,
tuned so that it can be directly applied at the physical masses
of the $b$ and $c$ quarks.
Our results are
$f_B     = 164 \err{14}{11} \pm 8$ MeV,
$f_{B_s} = 185 \err{13}{8} \pm 9$ MeV,
$f_D     = 194 \err{14}{10} \pm 10$ MeV, and
$f_{D_s} = 213 \err{14}{11} \pm 11$ MeV in the quenched approximation.
The first error in each case is statistical, and the second
is from perturbation theory.
We show that discretization errors are under control in our
approach, and smaller than our statistical errors.
The effects of the quenched approximation may raise our quenched 
result by up to 10\%.
\end{abstract}

\section{Introduction}

This paper uses lattice QCD to calculate 
the decay constants of $B$ and $D$ mesons 
using an $\calO (a)$ improved  action that takes
light quark actions as a starting point,
tuned so that it can be directly applied at the physical masses
of the $b$ and $c$ quarks.
The $B$ meson decay constant, $f_B$, is of particular phenomenological
interest, since it is responsible for most of the uncertainty
in current determinations of the Cabibbo-Kobayashi-Maskawa matrix
element $V_{td}$ from $B \overline{B}$ mixing.
Recent calculations of the decay constants of $B$ mesons give
much lower values than the earliest calculations 
done in the static approximation
because of several effects, all of which happened to be negative.
 Perturbative corrections turn out to be large and negative.
The static approximation has larger statistical errors than methods with
propagating quarks.  This led to contamination from excited
states, which raised the estimate of $f_B$.
 The use of smeared sources and propagating
quark methods mitigate this problem.
The $\calO (a)$  finite lattice spacing corrections likewise turned
out to be large and negative.

Conversely, the earliest results for lighter quarks 
 tended to be too low.
They were done using naive light quark methods with an incorrect and singular
quark field normalization which forced $f_M\sqrt{M}\rightarrow 0$
in the heavy quark limit.

The combination of a too high  $f_B \sqrt{M_B}$ from the static approximation
and a too low $f_D\sqrt{M_D}$ from light quark methods
 led to very large estimates of the $1/M$ corrections 
to $f_M\sqrt M$ in the static limit.
With more recent results, including those presented here,
the $1/M$ corrections are much reduced.
Refs.~\cite{Ber94} and \cite{Eic91} contain reviews of some of the early
work.

 Currently convenient inverse lattice spacings are no larger than
$a^{-1} \approx$ 2 or 3 GeV.  Therefore,
discretization errors that go like the quark mass in lattice units,
$m a$, to a power are
unpleasantly large for the $c$ quark and completely out of control
for the $b$ quark.
This means that standard light quark formulations for lattice fermion
actions cannot be used unaltered for the $b$ quark.

There are several ways of approximating heavy quarks in 
lattice QCD calculations with control over discretization errors.
These include Nonrelativistic QCD \cite{Cas86,Lep87,Lep92}  (NRQCD),
the static approximation \cite{Eic87,Hil90}, 
and the approach of this paper, described  in Refs.~\cite{Kha97,K&M93}.
The methods vary significantly both in ease of application and 
in suitability for various calculations.
The last takes the light quark actions of  Wilson \cite{Wil74} 
and Sheikholeslami and Wohlert (SW) \cite{She85}
 as its leading approximation
but adds correction operators that end up resembling those
of NRQCD rather than those of the standard Symanzik improvement
program.
Sec.~\ref{methods} contains a general discussion of the various
methods for heavy quarks.
It comments on   our bettered understanding  about which
methods work best in which situations,
not all of which was expected in advance.
Sec.~\ref{action} treats in detail the particular form of the
action used in this paper.
Our numerical results are presented in Sec.~\ref{results},
and compared with other recent results in Sec.~\ref{comp}.

\section{ Methods for Heavy Quarks on the Lattice}\label{methods}

This work uses a formalism for propagating heavy quarks that
reduces to the ordinary light quark formalism in the light quark 
limit~\cite{Kha97,K&M93}.
We begin by comparing the various lattice approaches to heavy quarks.
NRQCD is based on an expansion in nonrelativistic operators
(rotationally invariant but not  Lorentz invariant)
similar to that used in calculating relativistic corrections 
in the hydrogen atom.
It can be thought of as arising from a discretization of the 
action arising after a Foldy-Wouthuysen-Tani (FWT) transformation
of the quark fields:
\begin{equation}\label{FWT}
\psi \rightarrow \exp(\theta D_i \gamma_i ) \psi,
\end{equation}
with $\theta$ chosen so that
\begin{eqnarray}\
D \! \! \! \! /+m &\rightarrow&  D_0	\gamma_0 +m 
	\nonumber \\ \label{NR}
  &-&  	 \frac{\bD^2}{2m} - \frac{(\bD^2)^2}{8m^3} + \ldots .
\end{eqnarray}
The rest mass term does not affect the dynamics of nonrelativistic
particles and is conventionally removed.
Increasing accuracy is achieved by truncating the series with 
increasing numbers of terms.

In $B$ physics, one can use the simplest of all the methods,
the static approximation,  which is the truncation of the preceding
series to a single term.
Then the heavy quark propagator is a simple Wilson
line in the time direction.
It is  most useful for the heaviest quarks.
It is not much used recently because it has a much worse 
signal to noise ratio than methods using propagating quarks, 
which is clear in retrospect but was not foreseen.

The third method, used in this paper, can be thought of as arising from
 a partial FWT transformation:
\begin{equation}
\psi \rightarrow \exp(\theta' D_i \gamma_i ) \psi,
\end{equation}
and
\begin{eqnarray}
D \! \! \! \! /+m &\rightarrow&  D_0	\gamma_0 +m + a_1  D_i \gamma_i 
	\nonumber \\
  &-&  a_2 \left( \frac{\bD^2}{2m}  + \ldots \right),
\label{partialFTW}
\end{eqnarray}
where $\theta'< \theta$.
This appears to be a crazy thing to do, producing an action that
combines the  defects of the transformed and the untransformed 
actions.
On the other hand, it turns out that this is the action we 
have been using for a long time.
The Wilson and SW actions have just this form.
In addition to the usual  $\psibar D \! \! \! \! /  \psi$ term,
the $\psibar D^2 \psi$ term added to cure the doubling problem 
also contributes to the kinetic energy, as in Eqn.~\ref{partialFTW}.
The relative strengths of the two terms change as $m a$ is increased.
 
For the Wilson action, $a_1=1$, and $a_2=  m a$.
These have the property that the nonrelativistic
$\psibar D^2 \psi$ term takes over automatically from
the Dirac-style kinetic energy term $\psibar D \! \! \! \! /  \psi$
 as $ma \rightarrow \infty$
in the Wilson action.
The Wilson action turns into a nonrelativistic action
in the large mass limit.
The heavy quarks in heavy-light mesons are highly nonrelativistic.
($p/m_{ch} \sim \Lambda_{\rm QCD}/m_{ch} \sim.2$ and $p/m_b\sim .06$.)
Therefore, it is the Wilson term, rather
than the Dirac term, that contributes most to the
heavy quark kinetic energy.

 We can write a lattice energy-momentum relation
\begin{equation}\label{energy-expansion}
E^2 = M_1^2 + \frac{M_1}{M_2}{p}^2 + \ldots,
\end{equation}
where the ``rest mass''
\begin{equation}\label{M1}
M_1=E({0}),
\end{equation}
and the ``kinetic mass''
\begin{equation}\label{M2}
M_2^{-1}=(\partial^2E/\partial p_i^2)_{{\scriptstyle p}={0}}.
\end{equation}
The Wilson action and most other actions have
 the  property that
   $M_1$ does not equal  $M_2$ for $ma\ne 0$.
For nonrelativistic particles, the rest mass does not affect the dynamics
and the kinetic mass governs the leading important term.
Therefore, the rest mass is normally simply omitted from the action 
in NRQCD and in the static approximation, 
although there is no harm (and no benefit) in including it.
However, it cannot be set to zero  if one wishes to recover
a sensible light quark limit.
It is easy to find a Wilson style action which does satisfy 
$M_{1} = M_{2}$
by letting the hopping parameter for the time direction, $\kappa_t$,
differ from the one in the spatial directions, $\kappa_s$.
 The two $\kappa$'s can be separately tuned so that
$M_1=M_2$  with no loss of predictive power,
by requiring Lorentz invariance in 
Eqn.~\ref{energy-expansion}.
Nevertheless, since $M_1$ doesn't affect the physics of nonrelativistic
systems, the Wilson and SW actions can be used without problems 
for nonrelativistic systems as long as the kinetic mass, and not the rest mass,
 is used to set the quark mass. We can use $\kappa_t=\kappa_s$
with an incorrect $M_1$ and a correct $M_2$.

The Wilson and SW actions and the action of NRQCD all
use one-hop time derivatives.
When the action includes a large rest mass, 
this is a requirement, since two-hop
time derivatives introduce new  states with complex energies when $ma>1$.
Therefore, heavy-quark improvements of the Wilson action cannot
follow the conventional Symanzik program of adding two-hop space-like
and time-like
interactions at ${\cal O}(a^2)$, but must follow  NRQCD 
and correct only the spatial interactions.
The existence of  the transfer matrix and the Hamiltonian ensures that
this is possible.

The parameters of the action in our approach must have nontrivial
mass dependence, just as  those of NRQCD do.
For $ma>1$, the wave function normalization, the relation between
the physical mass and the hopping parameter, etc., are completely different
from their $ma=0$ values.
When $ma<1$,  this mass dependence may be expanded
in power series.
For the Wilson and SW Lagrangians,
this yields just the usual series of operators, with the same
coefficients.
If the mass dependent coefficients in our style of 
interpreting the Lagrangian,
\begin{eqnarray} 	\label{lagofm}
 {\cal L}& = &  m(am) \psibar\psi
+z(am) \psibar D \! \! \! \! /  \psi
   + c(am) \psibar \sigma_{\mu\nu} F_{\mu\nu} \psi +\ldots,
   \end{eqnarray}
are expanded, the usual series of powers of $ma$ multiplying identical
operators must result:
\begin{eqnarray} 	\nonumber
 {\cal L}& = &  m \psibar\psi + a_1 m^2 a\psibar\psi + \ldots\\ \nonumber
&+&z \psibar D \! \! \! \! /  \psi+ z_1 ma \psibar  D \! \! \! \! / \psi 
   + z_2 m^2 a^2  \psibar D \! \! \! \! / \psi+ \ldots \\       
&+& c_{SW} \psibar \sigma_{\mu\nu} F_{\mu\nu} \psi + 
   c_{2} ma \psibar \sigma_{\mu\nu} F_{\mu\nu} \psi + \ldots .
\end{eqnarray}
Sometimes in the Symanzik program,
$\psibar \sigma_{\mu\nu} F_{\mu\nu} \psi $ and
 $ ma \psibar \sigma_{\mu\nu} F_{\mu\nu} \psi$  are  spoken
of  as if they were totally unrelated operators.
 Even for light quarks, it makes more sense to think of them 
as different terms in an $ma$ expansion of an $ma$ dependent coefficient
analogous to the expansion in $\alpha_s$ of an $\alpha_s$
dependent coefficient.
The fact that coefficients in the usual approach blow up as
$ma\rightarrow \infty$
is a property of the expansion and not of the required functions
in Eqn. \ref{lagofm} themselves.
They stay well behaved if sensible normalization conditions are applied.

With the added ingredient of decoupling the timelike and spacelike
parts of operators, it becomes possible to formulate an
action that is systematically improvable, even for large $ma$.
While for $ma>1$ the action becomes very similar to NRQCD
in its behavior, for $ma<1$ it may be regarded as a
resummation of the usual operators of the Wilson and SW actions
 to all orders in $m$.

For physics involving the $b$ quark,
NRQCD methods are often easier and therefore more accurate.
Taking the $ma\ll 1$ limit is not possible, but correction operators
are not too hard to organize and add. 
This is particularly important in the $\Upsilon$ system,
where momenta are rather large, and correction operators are important
in obtaining such things as the correct spin-dependent spectrum.
In $B$ mesons, on the other hand, the $b$ quarks are extremely 
nonrelativistic. 
$(v/c)^2 \sim (0.3\ {\rm GeV}/ 5.0\ {\rm GeV})^2 \sim 0.3\%$.
Therefore, only the first few operators in the action,
which are the same in the two methods, are important.
As long as these are normalized in the same way,
it matters little what additional operators are hanging around.
Therefore, in $B$ mesons the effects of the differences between
NRQCD and our heavy quark methods are minimal, and
the two methods should yield nearly identical results.

For physics involving the $c$ quark,
heavy quark methods that can recover a relativistic form of the action
are often easier and more accurate.
The series in $v^2$ is less convergent, so eliminating the need for it
by taking $a$ toward zero while recovering  a relativistic action
may be more convenient.
For hadrons containing charmed quarks, it is  possible to do
calculations with the Wilson and SW actions even with the old interpretation
of the coefficients.
However, since $m_{ch}a \approx 5 a \Lambda_{QCD}$, our ability to
sum up the required series in $ma$ exactly 
is likely to produce a faster approach to the continuum limit
than naive light quark methods.

For physics involving either the $b$ or $c$ quarks,
both NRQCD and our method can be used successfully.
 Even when one works better or worse than the other, we still learn something.

\section{The Action Used in this Calculation}\label{action}
We have used the approach for heavy quarks outlined above
 to calculate the decay constants
of the $D$, $D_s$, $B$, and $B_s$ mesons.
We start with an action corrected to $\calO (a)$, which in general is
\begin{eqnarray}\label{kappa-kappa}
S &=& {\displaystyle \sum_n} \bar{\psi}_n\psi_n  	\nonumber \\
&-& \kappa_t {\displaystyle \sum_{n}} \left[\rule{0.0em}{0.97em}
\bar{\psi}_n(1-\gamma_0)U_{n,0}\psi_{n+\hat{0}}
+
\bar{\psi}_{n+\hat{0}}(1+\gamma_0)U^\dagger_{n,0}\psi_n
\right] \nonumber \\[1.0em] 
&-&\,\kappa_s {\displaystyle \sum_{n,i}} \left[\rule{0.0em}{0.97em}
\bar{\psi}_n(1-\gamma_i)U_{n,i}\psi_{n+\hat{\imath}}
+
\bar{\psi}_{n+\hat{\imath}}(1+\gamma_i)U_{n,i}^\dagger\psi_n
\right]	\nonumber \\
&+& \ihalf c_B \kappa_s \sum_{n;i,j,k}\varepsilon_{ijk}
\bar{\psi}_n\sigma_{ij}B_{n;k}\psi_n 		\nonumber \\
&+& ic_E \kappa_s \sum_{n;i}
\bar{\psi}_n\sigma_{0i}E_{n;i}\psi_n.
\end{eqnarray}
The second and third terms are the timelike and spacelike
pieces of the kinetic energy of the Wilson action.
The fourth and fifth terms are the spacelike and timelike pieces of
the SW correction operator.

The action used to calculate $f_M\sqrt{M}$ may be simplified from this form.
As previously noted, 
when $ma\ne 0$, we find that $M_1  \ne M_2$ and that higher
orders in $p$ in Eqn.~\ref{energy-expansion}
do not vanish as they should.
Explicitly, we define
\begin{equation}\label{mass}
m a =\frac{1}{2\kappa_t} - [1+3\zeta],
\end{equation}
where
$\zeta \equiv \kappa_s/\kappa_t$.
Then we can calculate the free lattice propagator and expand in $pa$ to derive
\begin{equation}\label{M1-tree}
M_1 a= \log(1 + m a),
\end{equation}
and
\begin{equation} \label{M2-tree}
\frac{1}{M_2a}=\frac{2\zeta^2}{ma(2+ma)}+\frac{\zeta}{1+ma}.
\end{equation}
For nonrelativistic fermions, the 
  physics of the system is influenced hardly at
all by the rest mass $M_1$.
 $M_2$ controls the leading order dynamics.
We can therefore  set 
\begin{equation}
\kappa_t=\kappa_s=\kappa,
\end{equation}
but we must tune $\kappa$ so that $M_2$ equals the physical quark mass.

Likewise, the physics of heavy-light mesons is 
insensitive to the final, spin-orbit term in Eqn.~\ref{kappa-kappa}.
(This not true of some other quantities in heavy quark physics,
such as the fine structure of quarkonia.)
We can therefore set
\begin{equation}
c_E=c_B=c,
\end{equation}
and tune $c$ to get the physics of the fourth operator in 
Eqn.~\ref{kappa-kappa}, the $\mbf \sigma \cdot B$ operator,
correct.
At  tree level,
this requires the same value that it has for light quarks in
the SW action, $c=c_{SW}=1$.

The decay constant~$f_M$ parametrizes the matrix element 
\begin{equation}
\langle 0|A_\mu|M,{p}\rangle=p_\mu f_M
\label{fB}
\end{equation}
of the axial vector current
between the pseudoscalar meson~$M$ and the vacuum.
Here the state $|M,{p}\rangle$ has the standard, relativistic 
normalization.
Like the action, the axial vector current~$A_\mu$ must be specified 
through ${\cal O}(a)$.
At tree level we take
\begin{equation}
A_\mu=2\sqrt{\kappa_{t,h}\kappa_{t,l}}Z_A\bar{\Psi}_h\gamma_\mu\gamma_5\Psi_l,
\label{axial current}
\end{equation}
where the subscripts $h$ and $l$ denote the heavy and light quarks,
$Z_A$ is a mass-dependent (re)normalization factor, 
and $\Psi$ denotes a (rotated) field, specified below.

The factors $\sqrt{2\kappa}$ arise from naive quark wave function
mass dependence in long historical use.
Wilson used the field normalization 
\begin{equation}\label{znaive}
z_{\rm wf}^{\rm naive} = \sqrt{2 \kappa}
\end{equation}
for massless fermions~\cite{Wil74}. 
It is easy to see  that this is correct only in
the $ma=0$ limit that Wilson was considering.
Away from $ma=0$, 
a straightforward examination of the free propagator shows that,
if the quarks are to have their conventional canonical normalizations,
the correct normalization is
\begin{equation}\label{zw}
z_{\rm wf} = \sqrt{1-6\kappa}
\end{equation}
 for the Wilson and SW actions, or 
\begin{equation}\label{zus}
z_{\rm wf} = \sqrt{1-6\kappa_s}
\end{equation}
if $\kappa_t \ne \kappa_s$.
These are equal to the conventional but incorrect normalization 
for $ma=0$:
$\sqrt{2\kappa}=\sqrt{1-6 \kappa}=1/2$.
As $m_ha\to\infty$ ($\kappa_{t,h}\to 0$), 
the naive normalizations force~$f_M$ incorrectly 
to zero, unless the mass dependence of~$Z_A$ compensates accordingly.
Indeed, the early estimates of $f_D$ that were too low stem from the 
naive assumption $Z_A\approx1$, even when $m_ha\sim 1$.
Combining Eqns.~\ref{znaive} and \ref{zus}, we see that
\begin{eqnarray}\label{eq:Z_A}
	Z_A &=&	\sqrt{
	\frac{1-6\kappa_{s,h}}{2\kappa_{t,h}}
	\frac{1-6\kappa_{s,l}}{2\kappa_{t,l}}
			}\\
		&=&	\sqrt{(1+m_ha)(1+m_la)},
\end{eqnarray}
where the second expression follows from Eqn.~\ref{mass}.

The remaining ingredient of Eqn.~\ref{axial current} is the rotated 
field
\begin{equation}
\Psi=(1+d_1\ {\gamma_i}{D_i})\psi,
\end{equation}
 where $i$ is summed over the three spatial directions
and, at tree level,
\begin{equation}
d_1=\frac{\zeta(1+ma)}{ma(2+ma)}-\frac{1}{2M_2a}.
\end{equation}
The second term of $d_1$ is the same as the $1/m$ correction 
present in NRQCD.
It is a consequence of the Foldy-Wouthuysen-Tani transformation in 
Eqn.~\ref{FWT}.
The first term of $d_1$ allows for the fact that the implicit 
Foldy-Wouthuysen-Tani transformation of Wilson-like actions takes 
$\theta'\neq\theta$.
The numerical value of $d_1$ depends on the tuning of 
$\zeta=\kappa_s/\kappa_t$.
In most cases, and in particular when $\zeta=1$, the contribution to 
$f_M$ proportional to $d_1$ is ${\cal O}(m\Lambda a^2)$ when $ma\ll 1$,
but ${\cal O}(\Lambda/m)$ when $ma\gg 1$.
In the latter region it is essential to include it, if one is 
interested in the $1/m_h$ corrections to the static limit.
It also makes sense to include the rotation when $ma< 1$.

The dominant mass-dependence exhibited in Eqn. \ref{eq:Z_A} 
persists beyond tree level~\cite{Kha97}.
Quantum effects require further terms to be added to the right-hand side 
of Eqn.~\ref{axial current} to construct an
${\cal O}(a)$-improved axial vector current.
In our numerical work we neglect them and we set $d_1$ to its 
tree-level, mean-field improved value.

Thus, to $\calO (a^0) $ and $\calO (a^1)$, we
can use  precisely the Wilson 
and the SW actions for much heavy quark physics.
 In this paper we will use the SW action:
\begin{eqnarray}\label{SSW}
S &=& {\displaystyle \sum_n} \bar{\psi}_n\psi_n  	\nonumber \\
&-&\,\kappa {\displaystyle \sum_{n,\mu}} \left[\rule{0.0em}{0.97em}
\bar{\psi}_n(1-\gamma_\mu)U_{n,\mu}\psi_{n+\hat{\imath}}
+
\bar{\psi}_{n+\hat{\imath}}(1+\gamma_\mu)U_{n,\mu}^\dagger\psi_n
\right]	\nonumber \\
&+& \ihalf c_{SW} \kappa \sum_{n;\mu\nu}
\bar{\psi}_n\sigma_{\mu\nu}F_{n;\mu\nu}\psi_n.
\end{eqnarray}
There are two key differences between
 our use of this action for heavy quarks and the naive approach.
\begin{enumerate}
\item The use of the correctly normalized quark fields and currents,
 and the additional three dimensional field rotation for the heavy quark
in the  currents.
\item The use of the kinetic mass rather than the rest mass
to set the quark mass.
\end{enumerate}
At higher orders in $a$, it is no longer true that ordinary forms
of the light quark action may be used for heavy quarks.
Two-hop corrections for
 time derivatives cannot be used since they introduce new
states with energies which become complex when $ma>1$.
$\calO (a^2)$ corrections must have a nonrelativistic form as in NRQCD.

Having determined the coefficients in the action at tree level,
we now discuss their renormalization.
In general, the coefficients $c_n$ of operators $O_n$ in a Wilsonian effective
action depend on the physical parameters of the theory and on the
renormalization scale.
For lattice QCD we can write
\begin{equation}
S(m/\Lambda_{QCD}, \Lambda_{QCD}\ a) = \sum_n c_n(m a, \alpha_s) O_n,
\end{equation}
where $m$ and $\alpha_s$ are the bare parameters of the lattice theory.
In the usual Symanzik approach, these coefficients are 
expanded both in $\alpha_s$ and in $ma$.
It is usually an advantage if an expansion in $\alpha_s$ can be avoided.
Likewise, as remarked earlier, there is an advantage in not
expanding in $ma$.
The expansion in $ma$ is the source of the
breakdown of the method for $ma>1$.  If the $c_n$ are not expanded
in $ma$, applying standard normalization conditions
 yields well behaved $c_n$ for large $ma$, as shown at
tree level for a few important cases above.
It is of course necessary that this is true for the $c_n$ in the
full quantum theory as well.

Most quantum corrections done so far have only been calculated
in perturbation theory.
It is clear that the tadpoles, which dominate many lattice
perturbation theory calculations, 
create well-behaved contributions to the renormalization of
these actions at all quark masses.
Tadpole-improved perturbation theory \cite{Lep93} 
suggests that the dominant perturbative correction to the
renormalization of coefficients of operators is given by 
short distance tadpoles.
They can be estimated by noting that
the link 
$U_\mu=1+i g_0A_\mu-\frac{1}{2} g_0^2A_\mu^2+\ldots$
does not fluctuate around 1, as is implicitly assumed in the
usual perturbative expansion,
but around some smaller value, $u_0$.
The so-called mean link
$u_0$ can be estimated from the expectation value of any link-containing
operator such as the plaquette ($\langle U_P \rangle  = u_0^4$)
or the expectation value of $U_\mu$ itself in Landau gauge.
By counting links in the actions 
Eqn.~\ref{kappa-kappa} or Eqn.~\ref{SSW}
one can see that the usual tadpole estimates of corrections to
operator coefficients
$\tilde{\kappa} = \kappa u_0$
and
$\tilde{c}_{SW} = c_{SW} u_0^3$
are still valid for heavy quarks even though the one-loop
coefficients differ slightly.

The one-loop coefficients which have been calculated so far have the
required property that they stay small and well-behaved
 as $ma$ is varied from 
0 to $\infty$~\cite{Mer94,Kur97}.
The part of the one-loop correction to the local  
heavy-light axial vector current arising from the leading operators
in the SW action  has been calculated by Aoki et al.~\cite{Has97}.
It also is well-behaved for all quark masses,
and has been incorporated into our results.
The part of the correction arising from the
$\calO (a)$ correction to the axial vector current
 is currently being calculated~\cite{Has97b}.

\begin{figure}
\epsfxsize= .85\textwidth
\epsfbox{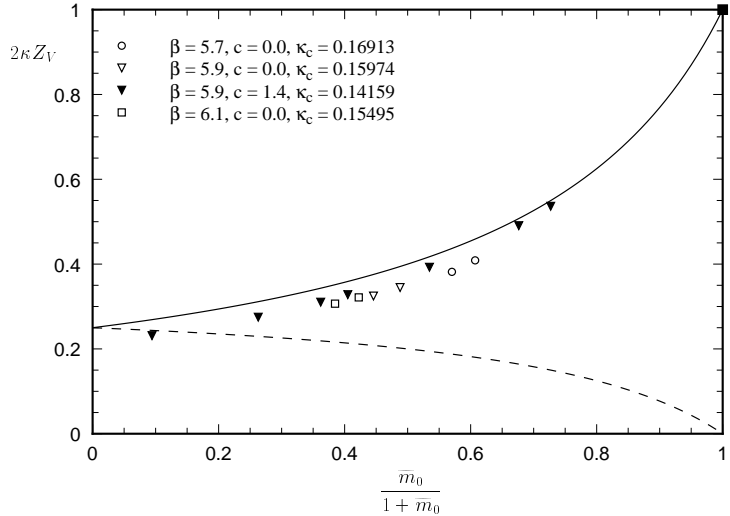}
\caption[]{The current normalization factor $2\kappa Z_V$
calculated nonperturbatively from the quantity
$\langle J/\psi|V_4|   J/\psi    \rangle$, where 
 the (unnormalized) local charge $V_4 = \overline{\psi}\gamma_0 \psi$
should simply count the numbers of fermions when properly normalized.
The upper curve is $1-6\tilde{\kappa}$, from mean-field improving
Eqn.~\ref{zw}.
The lower curve is the naive ansatz $2\tilde{\kappa}$.
} \label{fig0}
\end{figure}

One important quantity can be easily computed nonperturbatively,
to see that the required good behavior persists to all orders 
of perturbation theory.
This is the quark wave function renormalization, 
which is required for all current renormalizations.
This is part of the vector current, which is easy to normalize
nonperturbatively by inserting the charge into any physical state,
such as a 1S quarkonium state~\cite{Kha97}.
Fig.~\ref{fig0} shows the result of such a calculation,
compared with the tadpole improved tree-level correct normalization
and the naive normalization.
The vector current normalization factor $Z_V$ is defined analogously
to the axial vector normalization factor $Z_A$ in Eqn.~\ref{eq:Z_A}.
It is easy to see that the good behavior of the correct normalization
in Eqn.~\ref{zw} is preserved when full quantum effects are
included nonperturbatively.

\section{Numerical Results}\label{results}

\begin{table}[tbh]
\begin{center}
\begin{tabular}{cccc}
\hline
\hline
{\rule[-3mm]{0mm}{8mm}$\beta$} & 6.1 & 5.9 & 5.7 \\
\hline
 Volume                       & $24\times 48$ & $16\times 32$ & $12\times 24$ \\
\hline
Configurations                  & 100 & 350 & 300  \\
\hline
$c_{\rm SW}$                & 1.46 & 1.50 & 1.57 \\
\hline
$\alpha_V ( 1/a)$     & 0.222 & 0.259 & 0.330 \\
\hline
$\alpha_V ( 2/a)$     & 0.171 & 0.192 & 0.227 \\
\hline
$\alpha_V ( \pi /a) $ & 0.149 & 0.164 & 0.189 \\
\hline
$u_0$			    &0.8816        & 0.8734        & 0.8608 \\
\hline
$\kappa_l$                  & 0.1372 & 0.1382 & 0.1405 \\
		            & 0.1373 & 0.1385 & 0.1410 \\
			    & 0.1376 & 0.1388 & 0.1415 \\
			    & 0.1379 & 0.1391 & 0.1419 \\
			    &        & 0.1394 &		\\
\hline
$\kappa_h$                  &0.050   &  0.050 & 0.050 \\
			    & 0.099  & 0.093  & 0.070 \\
			    & 0.126  & 0.1227 & 0.089 \\
			    & 	     & 0.126  & 0.110 \\
			    &        &        &0.119 \\
\hline
$a^{-1}(f_K)$ (GeV)	    &2.21\er{7}{3} & 1.57\er{3}{2} &1.01\er{2}{1}\\
\hline
$a^{-1}(1P-1S)$	(GeV)	    &2.62\er{8}{9} & 1.80\er{5}{5} &1.16\er{3}{3} \\
\hline
\hline
\end{tabular}

\caption{Parameters used in the numerical calculations.}
\label{tab:details}
\end{center}
\end{table}

We turn now to a discussion of our numerical results.
A preliminary version of our results appeared in Ref.~\cite{Rya97}.
In Table~\ref{tab:details} we show some details of our calculations.
The light quark propagators are the same ones used to determine the
light quark masses in Ref~\cite{Gou96}.
They were calculated with the SW action using a mean-field improved
 coefficient $c_{SW}$.
$\alpha_V ( 1/a)$, $\alpha_V ( 2/a)$, and $\alpha_V ( \pi /a)$,
defined as in Ref.~\cite{Lep93},
are values of the strong coupling constant at several relevant scales.
$u_0$ is the mean link used in tadpole improvement.
$\kappa_l$ and $\kappa_h$ are the hopping parameters used
for the light and heavy quarks, respectively.
$a^{-1}(f_K)$ and $a^{-1}$(1P-1S) are the lattice spacings in
physical units as determined by $f_K$ and the 1P--1S splitting
of charmonium, respectively.
For $\beta=$ 6.1, 5.9, and 5.7, the hopping parameters closest to
the $b$ hopping parameters are $\kappa_b\approx$ 0.099, 0.093, and 
0.089 respectively.  Those closest to charm are 
$\kappa_{ch}\approx$ 0.126, 0.1227, and 0.119 respectively.
The hopping parameters
 closest to strange are 0.1373, 0.1385, and 0.1405
respectively.
Only modest interpolations in quark mass are necessary for
$B_s$ and $D_s$ mesons.
$B$ and $D$ physics must be obtained by chiral extrapolation to
the physical light mass limit.
Charm and bottom $\kappa$'s were obtained 
by demanding that spin-averaged kinetic masses of the
1S charmonium and bottomium states match experiment.
For the strange quark, the hopping parameter was fixed from the kaon mass.
We have also done calculations with static propagators for the heavy quarks,
which corresponds in our notation to $\kappa_s=0$ and $\kappa_t=1/2$.

\begin{figure}[tbh]
\epsfxsize=0.65\textwidth
\begin{center}
\leavevmode
\epsfbox{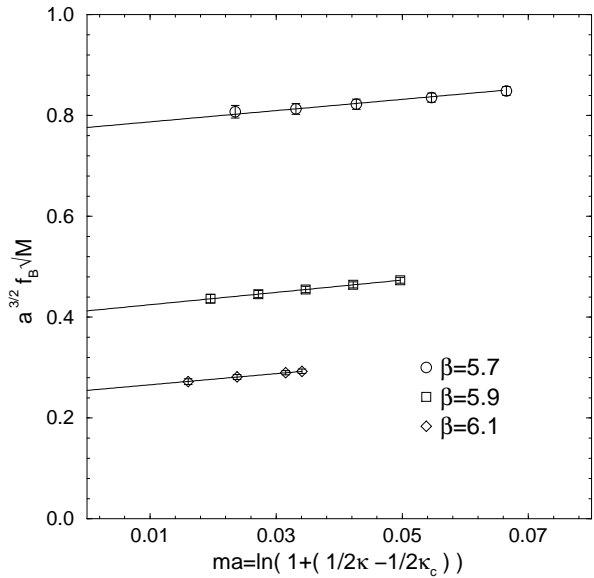}
\end{center}
\caption[]{
 Chiral extrapolations of $f_B \sqrt{M_B}$ to $m=0$
at $\beta=5.7$ (top),
$\beta=5.9$ (center), and $\beta=6.1$ (bottom).
} \label{chiral}
\end{figure}

Masses and matrix elements were obtained from minimizing $\chi^2$'s
using the the full correlation matrix.
Decay constants were obtained by dividing  matrix elements
by the square root of the experimentally measured meson mass.
Statistical errors were calculated with the bootstrap method,
using 1000 bootstrap samples in each fit.
Chiral extrapolations of $f_B \sqrt{M_B}$ at $\beta=$ 6.1, 5.9, and 5.7
are shown in Fig.~\ref{chiral}.
No problems with $\chi^2$ using linear fits were observed with the
$\kappa_l$'s shown.
In runs at lower quark masses, which we did not use here, 
occasional exceptional configurations
and poor $\chi^2$'s began to appear.

\begin{figure}[tbh]
\epsfxsize=0.65 \textwidth
\begin{center}
\leavevmode
\epsfbox{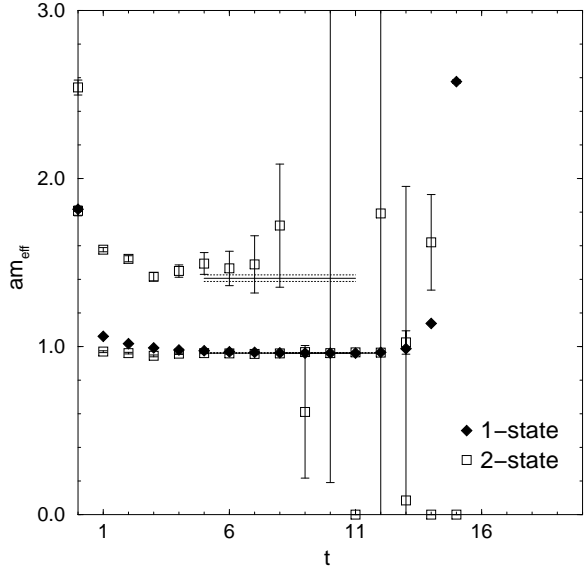}
\end{center}
\caption[]{
Comparison of results of one- and two-state fits at $\beta=5.9$,
$\kappa_h = 0.1227$ (near $\kappa_{ch}$), and
$\kappa_l =0.1385 $ (near $\kappa_s$). 
Good consistency between the two is observed in the fit range
for the lightest state.
} \label{fits}
\end{figure}

The possibility of contamination of the results by excited states
was checked by testing for consistency of one-, two-, and 
three-state fits using $\delta$ function, 1S, and 2S sources at both
source and sink.   
In the sources,  one quark
is smeared with the supposed wave function and the other is a 
delta function.
Shapes of the 1S and 2S sources were taken from
lattice Coulomb gauge valence quark wave functions of the $B$ and $D$ mesons.
Fig.~\ref{fits} shows a comparison of one- and two-state fits.
The measured energy splitting between the ground and first excited
states and the observed agreement between  the ground state
energies obtained in one- and two-state fits suggests that the
contamination from excited states is small
in the region used for fitting.
The measured values of $f_M\sqrt{M}$ in one-state fits and two-state
fits agreed to within statistical errors.

\begin{figure}[p]
\epsfxsize=.6 \textwidth
\begin{center}
\leavevmode
\epsfbox{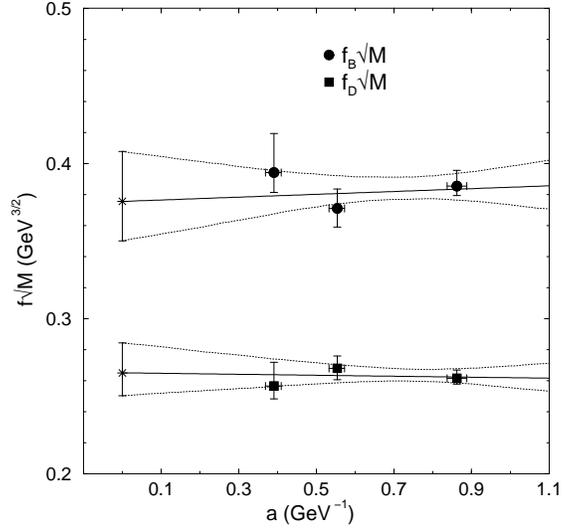}
\end{center}
\caption[]{
 $f_B \sqrt{M_{B}}$ and $f_D \sqrt{M_{D}}$ 
extrapolated to zero lattice spacing.
The statistical error bands are 1 $\sigma$ ranges of linear 
extrapolations found in bootstrap runs.
} \label{fig:fM}
\end{figure}

\begin{figure}[p]
\epsfxsize=.6 \textwidth
\begin{center}
\leavevmode
\epsfbox{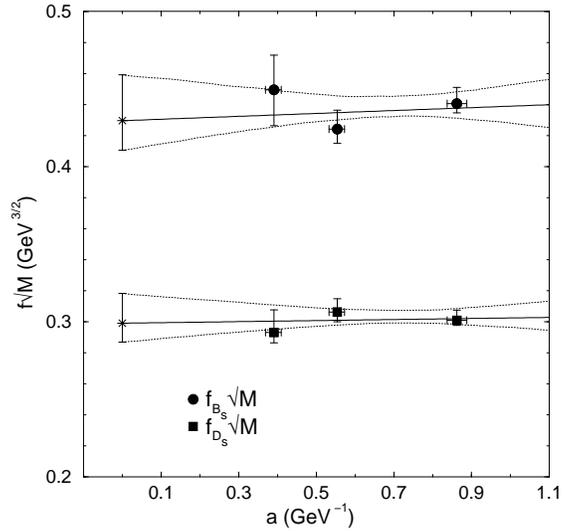}
\end{center}
\caption[]{
 $f_{B_s} \sqrt{M_{B_s}}$ and $f_{D_s} \sqrt{M_{D_s}}$ 
extrapolated to zero lattice spacing.
} \label{fig:fMs}
\end{figure}

The dominant finite lattice spacing errors for the Wilson action
are $\calO (a)$.
When these have been removed, as they have been in the SW action,
the remaining finite $a$ dependence arises from a combination of 
$\calO (\alpha_s^2)$, $\calO (\alpha_s a)$, and $\calO (a^2)$.
This means that attempts to extrapolate away any remaining 
finite $a$ dependence are uncertain, since we do not have a
theory of the functional form of the remaining dependence.
The ideal situation when using improved actions 
is negligible dependence of the results on $a$.
As shown in Figs.~\ref{fig:fM} and \ref{fig:fMs},
to within statistical errors, this has been achieved if
$f_K$ is used to determine the lattice spacing.
The results extrapolated to zero lattice spacing are consistent 
with the results at the smallest lattice spacings.
Thus, within our statistical errors we do not find any
evidence for discretization errors.

The least understood source of uncertainty is due to
the quenched approximation (the omission of sea quarks).
One effect of the quenched approximation is that different answers
for decay constants will be obtained depending on which 
physical quantity is used to set the lattice spacing in physical
units.
The appendix tabulates results using $f_K$, $f_\pi$ and
the 1P--1S splitting in charmonium to set the scale.
The well-measured physical quantity which most resembles the
heavy-light decay constants is $f_K$.
It therefore offers the best chance that some 
 statistical, systematic, and especially quenching
errors in $f_M$ will cancel out.
It is relatively easy to determine numerically.
($f_\pi$, for example, requires an additional chiral extrapolation
which introduces larger statistical error and unreliability.
Our $K$ consists of two degenerate quarks of mass $(m_s+m_d)/2$.)
Taking $a^{-1}$ from $f_K$ 
has the  pragmatic defect that it yields $f_M$'s that lie
at the bottom of the range given by all of the standard methods
of setting the lattice spacing, so that removing the quenched approximation
is more likely to move the results up than down.
One way of estimating the effect of the quenched approximation 
is to examine the spread of values obtained by setting the lattice
spacing in various reasonable ways.
However, as shown in the appendix, quantities similar to $f_M$
tend to give similar results, while quantities more dissimilar,
such as quarkonia spectra, can give results quite a bit different.
It is not easy to quantify what is ``reasonable''.
The most quantitative estimate so far of estimating the effects of
quenching with actual unquenched calculations was given in Ref.~\cite{Ber97}.
It compares results for quenched and unquenched calculations 
done in similar ways.
They report quenching uncertainties of around
$_{-0}^{+5}\%$ for the $D$ and $D_s$ mesons, and
$_{-0}^{+10}\%$ for the $B$ and $B_s$.

We now summarize our results and uncertainties.
In Figs.~\ref{fig:fM} and \ref{fig:fMs} we show results for 
$f_D\sqrt{M_D}$, $f_{D_s}\sqrt{M_{D_s}}$, $f_B\sqrt{M_B}$, and $f_{B_s}\sqrt{M_{B_s}}$,
using $f_K$ to set the lattice spacing in GeV.
Our final results for the decay constants and their ratios are
shown in Table~\ref{tab:results}.
Statistical errors in the decay constants are obtained from
bootstraps over extrapolations to $a=0$ linearly in $a$.
The $\chi^2$'s refer to these extrapolations.
Statistical errors in the ratios are obtained from bootstraps
over the ratios of extrapolations thus obtained.
\begin{table}
\begin{center}
\begin{tabular}{ccc}
\hline
\hline
                  & {\rule[-3mm]{0mm}{8mm}Final Value }      & $\chi^2/d.o.f$ \\
\hline
$f_B$             & $164\err{14}{11}\pm 8$ MeV  & 1.32/1\\
$f_{B_s}$         & $185\err{13}{8}\pm 9$ MeV  & 2.06/1  \\
$f_D$             & $194\err{14}{10}\pm 10$ MeV  & 0.77/1 \\
$f_{D_s} $        & $213\err{14}{11}\pm 11$ MeV  & 1.01/1 \\
$f_D/f_B$         & 1.19\err{6}{6}    & \\
$f_{D_s}/f_{B_s}$ & 1.14\err{4}{3}    &\\
$f_{B_s}/f_B$     & 1.13\er{5}{4}      & \\
$f_{D_s}/f_D$     & 1.10\err{4}{3}  & \\
$(f_{B_s}/f_B)/(f_{D_s}/f_D)$  	& 1.03\err{4}{4}	&	\\
\hline
\hline
\end{tabular}

\caption{Final results for the decay constants and ratios
in the quenched approximation.
The first error for the decay constants is from statistics,
the second is systematics, mostly from perturbation theory.
An additional uncertainty of perhaps 10\% is present in the decay constants
due to the quenched approximation, as discussed in the text.
The error in decay constant ratios is statistical.
}
\label{tab:results}
\end{center}
\end{table}

A common way of estimating the uncertainties from perturbation theory
is to do the perturbation theory at a range of plausible scales,
say $1/a$ to $\pi/a$.
That method does not give a sensible result in this case.
The scale variation has a much larger effect at large $a$ than at small $a$.
When the results are extrapolated linearly to $a=0$, 
they give the same answers to within a per cent, even though
the perturbative effect being extrapolated is not at all linear in $a$.
This is clearly an underestimate of the perturbative uncertainties.
These are due to $\calO (\alpha_s^2)$ terms of order 3\% 
plus the remaining $\calO (\alpha_s)$ term which is currently being calculated.
Assuming that this term has roughly the same value that it has in NRQCD,
we have taken 5\% for the perturbative uncertainty.

The remaining uncertainties in the quenched approximation
are likely to be  smaller than the statistical and perturbative uncertainties.
We do not see evidence in Figs.~\ref{fig:fM} and \ref{fig:fMs}
 for finite lattice spacing effects within our statistical
errors.
We estimate that
the effects of uncertainties in the determination of the heavy quark mass
are about a per cent or two.
Finite volume errors are likewise expected to be small for $B$ and $D$
mesons on our lattice volumes.

Quenching uncertainty estimates of around 10\% are common,
based on ranges of results using different quantities to set
the lattice spacing.
The crude, but quantitative estimates of quenching effects in Ref.~\cite{Ber97}
are within the range of the less quantitative estimates.
If we assume that the effects of the quenched approximation are similar
for different quark actions, and apply the largest of their
 quenching uncertainties to our quenched results, we obtain,
$f_B     = 164 \err{14}{11} \pm 8  \err{16}{0}$  MeV,
$f_{B_s} = 185 \err{13}{8}  \pm 9  \err{18}{0}$  MeV,
$f_D     = 194 \err{14}{10} \pm 10 \err{20}{0}$  MeV, and
$f_{D_s} = 213 \err{14}{11} \pm 11 \err{21}{0}$  MeV,
where the uncertainties are statistical, systematic other than quenching,
and quenching.  The last uncertainty is less quantitative than the first two.

Perturbative errors cancel in ratios of decay constants.
It is possible, but not proven, that deviations of ratios from one have
quenching uncertainties of the usual 10-20\% size.
In that case they would be significantly smaller than our
statistical uncertainties in these quantities.
We have therefore quoted only statistical errors for these quantities.

Our value for $f_{D_s}$ is compatible with the world average
experimental value $f_{D_s}=241\pm21\pm30$ MeV \cite{Ric97}.
The double ratio
$(f_{B_s}/f_B)/(f_{D_s}/f_D)$ 
is within a few per cent of one, in accordance with expectations from
chiral symmetry and heavy quark symmetry~\cite{Gri93}.

\section{Comparison with Other Results} \label{comp}

 Refs.~\cite{Ono97,Arifa,Wit97,Ber97a} contain  recent reviews
of heavy-light decay constants.
Most recent calculations are consistent,
with values of around 160 MeV for $f_B$ and around 195 MeV for $f_D$.
The JLQCD collaboration has completed a calculation of 
heavy-light decay constants using  methods almost identical
to ours~\cite{Aok97}.
They have used the $\rho$ mass rather than $f_K$ to set the lattice spacing,
but that makes a negligible difference.
The results should agree to within statistical errors,
and they do.
They obtain $f_B=163\pm9\pm8\pm11$ MeV,
where the errors arise from statistics, systematics
other than quenching, and quenching.
These results are also reasonably compatible with results using 
Wilson fermions without the $\calO (a)$ correction
which rely on extrapolating linearly  to $a=0$ to remove the leading errors.
The MILC collaboration has obtained in preliminary results
$f_B=153\pm10\err{36}{13}\err{13}{0}$ MeV,
where the errors arise from statistics, systematics other than quenching,
 and quenching~\cite{Ber97}.
Allton et al. obtain higher results than the others,
with $f_B=180\pm 32$ and $f_D= 221\pm17$ MeV, though not inconsistent
within errors \cite{All97}.
They have used a tree level coefficient for the clover term which is 
2/3 the value after quantum corrections.
This should leave some residual discretization error, but they do not
try to extrapolate it away, and it is responsible for their large error 
estimate for $f_B$.
In addition, they have not performed  the $f_B$ calculation 
at the physical $b$ quark mass, but have extrapolated from the 
region $m_b a <1$.
This may add a source of uncertainty which is hard to quantify.

\begin{figure}[tbh]
\epsfxsize=.7 \textwidth
\begin{center}
\leavevmode
\epsfbox{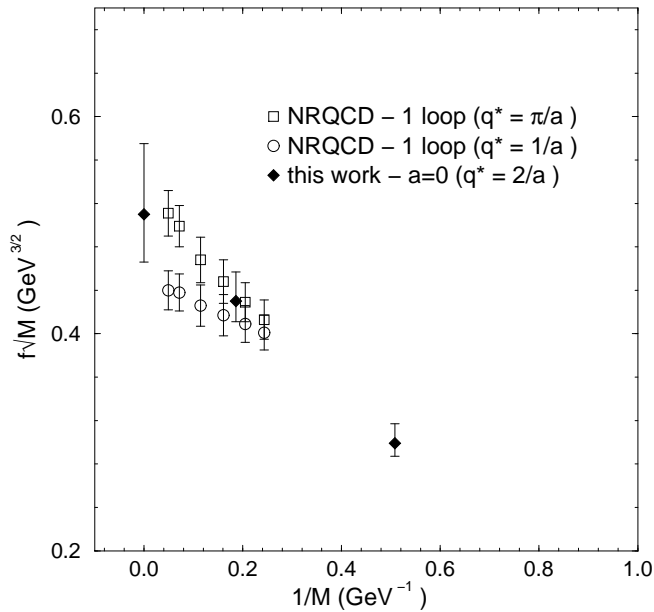}
\end{center}
\caption[]{
Comparison of the continuum limit results of this paper (filled symbols)
with the results of NRQCD (open symbols).
The errors shown are statistical.
} \label{fig:NRQCD}
\end{figure}

As remarked earlier, for the $B$ meson system,
the methods of NRQCD are also very similar to ours.
This is less true for the $D$, $\psi$, and $\Upsilon$ systems.
It is therefore interesting to compare recent results from the two methods.
Fig.~\ref{fig:NRQCD}
shows results for $f_M\sqrt{M}$ as a function of $1/M$ from this
paper and from a recent NRQCD calculation~\cite{Shi97}.
The NRQCD calculation includes a full $\calO (a)$ correction
and a full one-loop correction.
This paper uses a full $\calO (a)$ correction,
but is missing the part of the one-loop correction arising from the
 $\calO (a)$ correction to the current, which has not yet been calculated.
The one-loop corrected results are  close to each other
in the region of the $B$ mass, where we have reported our main results.
However, the NRQCD  one-loop corrections increase greatly 
for larger $M$, becoming as large as $-$25\% in the static limit,
the result of a  surprisingly large contribution from the $\calO (a)$
current correction.

For all of the heavy-light decay constants,
the largest source of uncertainty uncertainty, and the one under least
quantitative control, is the quenched approximation.
The uncertainties from all other sources of 
error  are now less than the approximately 10\% 
range usually associated with the quenched approximation.

\section*{Acknowledgments}

We thank S. Hashimoto and A. Ali Khan for communicating the results in
Refs.~\cite{Has97} and \cite{Shi97} to us prior to publication.
Fermilab is operated by Universities Research
Association, Inc. under contract with the U.S. Department of Energy.
AXK is supported in part through the DOE OJI program under grant 
no. DE-FG02-91ER40677 and through a fellowship from the Alfred P. 
Sloan foundation.

\appendix

\section{Appendix}
We present in Tables \ref{fBofa} -- \ref{fDsextrap}
 results for the decay constants obtained
with various fitting methods.
We give two tables for each of the four decay constants.
The first table contains results for each of the three lattice spacings.
 The second table gives the results of extrapolation to the continuum 
limit, along with the $\chi^2$ of the fit.
In each case, we performed the analysis 
with quark masses determined from  the 
spin-averaged 1S -onium state,
using one of five methods of determining the lattice spacing:
the 1P--1S charmonium splitting, $f_K$ from either
2- or 3-state fits, or $f_\pi$ from either 2- or 3-state fits.

\begin{table}[htpb] 
\begin{center}
\begin{tabular}{cccc}
\hline
\hline
{\rule[-3mm]{0mm}{8mm}$a^{-1}$} & $\beta=6.1$ & $\beta=5.9$ & $\beta=5.7$ \\
\hline
1P-1S             & 0.511\err{26}{35} & 0.455\err{19}{22} & 0.477\err{20}{22}\\
$f_\pi$ (2-state) & 0.419\err{37}{13} & 0.407\err{30}{13} & 0.440\err{25}{12} \\
$f_\pi$ (3-state) & 0.417\err{56}{38} & 0.363\err{21}{5} & 0.384\err{21}{12} \\
$f_K$ (2-state)   & 0.394\err{25}{13} & 0.371\err{12}{12} & 0.385\err{10}{6} \\
$f_K$ (3-state)  & 0.391\err{35}{10}  & 0.357\err{10}{10} & 0.387\err{10}{7} \\
\hline 
\hline
\end{tabular}
\caption[fBofa]{
$f_B \sqrt{M_B}$ in GeV$^\frac{3}{2}$
at $\beta =6.1$, 5.9, and 5.7, 
fitted with the lattice spacing
set in five different ways.
}
\label{fBofa}
\end{center}
\end{table}
\begin{table}[htpb]  
\begin{center}
\begin{tabular}{cccc}
\hline
\hline
{\rule[-3mm]{0mm}{8mm}$a^{-1} $} & $f_B\sqrt{M}(a= 0)$ & $f_B(a=0)$ & $\chi^2/d.o.f$ \\
\hline
1P-1S             & 0.490\err{42}{51} & 213\err{18}{22} & 2.04/1 \\
                   $f_\pi$ (2-state) & 0.386\err{58}{27} & 168\err{25}{12} & 0.51/1 \\
                   $f_\pi$ (3-state) & 0.349\err{63}{33} & 152\err{27}{14} & 1.63/1  \\
                   $f_K$ (2-state)   & 0.376\err{32}{25} & 164\err{14}{11} & 1.32/1  \\
                   $f_K$ (3-state)  & 0.331\err{34}{24} & 144\err{15}{10} & 3.51/1   \\
\hline
\hline
\end{tabular}
\caption[fBextrap]{$f_B \sqrt{M_B}$ in GeV$^\frac{3}{2}$ and 
$f_B$ in MeV extrapolated to $a=0$, 
fitted with the lattice spacing
set in five different ways.
}
\label{fBextrap}
\end{center}
\end{table}
\begin{table}[htpb]  \label{fBsofa}
\begin{center}
\begin{tabular}{cccc}
\hline
\hline
{\rule[-3mm]{0mm}{8mm}$a^{-1}$} & $\beta=6.1$ & $\beta=5.9$ & $\beta=5.7$ \\
\hline
1P-1S             & 0.582\err{26}{32} & 0.520\err{21}{21} & 0.545\err{22}{24}\\
$f_\pi$ (2-state) & 0.478\err{36}{10} & 0.465\err{33}{11} & 0.503\err{27}{13} \\
$f_\pi$ (3-state) & 0.476\err{60}{41} & 0.415\err{22}{2}  & 0.439\err{23}{12}\\
$f_K$ (2-state)   & 0.449\err{23}{10} & 0.424\err{12}{9}  & 0.440\err{11}{6}\\
$f_K$ (3-state)  & 0.445\err{32}{7}   & 0.408\err{8}{5}   & 0.442\err{10}{6} \\
\hline
\hline
\end{tabular}
\caption[fBsofa]{
$f_{B_s} \sqrt{M_{B_s}}$ in GeV$^\frac{3}{2}$
at $\beta =6.1$, 5.9, and 5.7, 
fitted with the lattice spacing
set in five different ways.
}
\end{center}
\end{table}
\begin{table}[htpb]
\begin{center}
\begin{tabular}{cccc}
\hline
\hline
{\rule[-3mm]{0mm}{8mm}$a^{-1} $} & $f_{B_s}\sqrt{M}(a= 0)$ & $f_{B_s}(a=0)$ & $\chi^2/d.o.f$ \\
\hline
1P-1S             & 0.567\err{42}{50} & 244\err{18}{22} & 2.80/1 \\
                   $f_\pi$ (2-state) & 0.444\err{61}{22} & 192\err{26}{10} & 0.59/1 \\
                   $f_\pi$ (3-state) & 0.395\err{68}{31} & 170\err{29}{13} & 1.81/1 \\
                   $f_K$ (2-state)   & 0.430\err{29}{19} & 185\err{13}{8} & 2.06/1 \\
                   $f_K$ (3-state)  & 0.374\err{31}{20} & 161\err{14}{9} & 6.51/1 \\
\hline
\hline
\end{tabular}
\caption[dum]{
$f_{B_s} \sqrt{M_{B_s}}$ in GeV$^\frac{3}{2}$ and 
$f_{B_s}$ in MeV extrapolated to $a=0$, 
fitted with the lattice spacing
set in five different ways.
}
\end{center}
\end{table}
\begin{table}[htpb]
\begin{center}
\begin{tabular}{cccc}
\hline
\hline
{\rule[-3mm]{0mm}{8mm}$a^{-1} $} & $\beta=6.1$ & $\beta=5.9$ & $\beta=5.7$ \\
\hline
1P-1S             & 0.332\err{16}{23} & 0.329\err{13}{15} & 0.323\err{13}{15} \\
$f_\pi$ (2-state) & 0.273\err{23}{9} & 0.294\err{20}{9} & 0.298\err{15}{8} \\
$f_\pi$ (3-state) & 0.271\err{35}{24} & 0.262\err{13}{2} & 0.261\err{13}{8} \\
$f_K$ (2-state)   & 0.256\err{15}{8} & 0.268\err{8}{7} & 0.261\err{5}{4} \\
$f_K$ (3-state)  & 0.254\err{21}{7} & 0.258\err{5}{4} & 0.262\err{5}{4} \\
\hline
\hline
\end{tabular}
\caption[dum]{
$f_D \sqrt{M_D}$ in GeV$^\frac{3}{2}$
at $\beta =6.1$, 5.9, and 5.7, 
fitted with the lattice spacing
set in five different ways.
}
\end{center}
\end{table}
\begin{table}[htpb] 
\begin{center}
\begin{tabular}{cccc}
\hline
\hline
{\rule[-3mm]{0mm}{8mm}$a^{-1} $} & $f_D\sqrt{M}(a= 0)$ & $f_D(a=0)$ & $\chi^2/d.o.f$ \\
\hline
1P-1S             & 0.339\err{27}{34} & 248\err{20}{24} & 0.03/1 \\
                   $f_\pi$ (2-state) & 0.262\err{38}{18} & 191\err{28}{13} & 0.43/1 \\ 
                   $f_\pi$ (3-state) & 0.268\err{39}{20} & 196\err{29}{15} & 0.06/1 \\
                   $f_K$ (2-state)   & 0.265\err{19}{14} & 194\err{14}{10} & 0.77/1 \\
                   $f_K$ (3-state)  & 0.249\err{15}{13} & 182\err{11}{10} & 0.01/1 \\
\hline
\hline
\end{tabular}
\caption[dum]{
$f_D \sqrt{M_D}$ in GeV$^\frac{3}{2}$ and 
$f_D$ in MeV extrapolated to $a=0$, 
fitted with the lattice spacing
set in five different ways.
}
\end{center}
\end{table}
\begin{table}[htpb]
\begin{center}
\begin{tabular}{cccc}
\hline
\hline
{\rule[-3mm]{0mm}{8mm}$a^{-1}$} & $\beta=6.1$ & $\beta=5.9$ & $\beta=5.7$ \\
\hline
1P-1S             & 0.380\err{16}{22} & 0.376\err{15}{15} & 0.372\err{15}{16} \\
$f_\pi$ (2-state) & 0.312\err{24}{8} & 0.336\err{24}{8} & 0.343\err{18}{9} \\
$f_\pi$ (3-state) & 0.310\err{40}{25} & 0.300\err{15}{1} & 0.300\err{16}{8} \\
$f_K$ (2-state)   & 0.293\err{15}{7} & 0.306\err{9}{6} & 0.301\err{7}{3}  \\
$f_K$ (3-state)  & 0.290\err{23}{5} & 0.295\err{5}{3} & 0.302\err{6}{3} \\
\hline
\hline
\end{tabular}
\caption[dum]{
$f_{D_s} \sqrt{M_{D_s}}$ in GeV$^\frac{3}{2}$
at $\beta =6.1$, 5.9, and 5.7, 
fitted with the lattice spacing
set in five different ways.
}
\end{center}
\end{table}
\begin{table}[htpb] 
\begin{center}
\begin{tabular}{cccc}
\hline
\hline
{\rule[-3mm]{0mm}{8mm}$a^{-1} $} & $f_{D_s}\sqrt{M}(a= 0)$ & $f_{D_s}(a=0)$ & $\chi^2/d.o.f$ \\
\hline
1P-1S             & 0.384\err{31}{35} & 274\err{22}{25} & 0.01/1 \\
                   $f_\pi$ (2-state) & 0.295\err{39}{16} & 210\err{28}{11} & 0.42/1 \\
                   $f_\pi$ (3-state) & 0.303\err{46}{20} & 216\err{33}{14} & 0.08/1 \\ 
                   $f_K$ (2-state)   & 0.299\err{19}{15} & 213\err{14}{11} & 1.01/1 \\
                   $f_K$ (3-state)   & 0.281\err{17}{11} & 201\err{12}{8} & 0.10/1 \\
\hline
\hline
\end{tabular}
\caption[fDsextrap]{
$f_{D_s} \sqrt{M_{D_s}}$ in GeV$^\frac{3}{2}$ and 
$f_{D_s}$ in MeV extrapolated to $a=0$, 
fitted with the lattice spacing
set in five different ways.

}
\label{fDsextrap}
\end{center}
\end{table}

\end{document}